\documentclass[preprint]{aastex}
\usepackage{graphicx}
\begin{document}
\title{Kick velocity induced by magnetic dipole and 
quadrupole radiation}
\author{Yasufumi Kojima$^*$ and Yugo E. Kato }
\affil{
Department of Physics, Hiroshima University, Higashi-Hiroshima 
739-8526, Japan }
\email{$^*$ kojima@theo.phys.sci.hiroshima-u.ac.jp}
\begin{abstract}
  We examine the recoil velocity induced by the superposition 
of the magnetic dipole and quadrupole radiation
from a pulsar/magnetar born with rapid rotation.
The resultant velocity depends on not the magnitude, but rather 
the ratio of the two moments and
their geometrical configuration.
The model does not necessarily lead to high spatial 
velocity for a magnetar with a strong magnetic field, which 
is consistent with the recent observational upper 
bound.  The maximum velocity predicted with this model is slightly
smaller than that of observed fast-moving pulsars.
\end{abstract}

  \section{INTRODUCTION}
  The surface magnetic field strength $B_{s}$ 
of a pulsar is conventionally estimated by
matching the rotational energy loss rate 
with the magnetic dipole radiation rate, that is,
$B_{s} \approx $
$(3 c^{3} I P \dot{P} )^{1/2} /(2\sqrt{2} \pi R_{s} ^{3})$,
where $I$ is the inertial moment, $R_{s}$ is the stellar radius,
$P$ is the spin period, and $\dot{P}$ is the time derivative 
of the spin period.
The precision of this approximation is only at the order of magnitude 
level because actual energy loss is not well described by
magnetic dipole radiation in a vacuum.
A more realistic model with current flows and radiation losses
is required, but has not yet been established. 
A simple estimate provides
$B_{s} \approx 10^{12}$G for typical radio and X-ray pulsars,
and $B_{s} \approx 10^{13}$-$ 10^{15}$G for magnetars, 
although the level of the approximation must be noted.
Dynamo action in a rapidly rotating proto-neutron star with
$P \approx 1 $ ms is proposed as a mechanism
for this amplification by 2-3 orders of magnitude 
(see e.g., \cite{ThDu93, BoReUr03, BoUrBe06}). 
Actual upper limit of $B_{s} $ generated in the convective proto-neutron 
star is estimated as $10^{15}-10^{16}$G, beyond which all sorts of 
instabilities are suppressed by strong magnetic fields
\citep{MiPoUr02}. 

  Recent numerical simulations of dynamo action can be used to study 
the large-scale fields in fully convective rotating stars.
For example, non-axisymmetric fields are generated
in the case of uniform rotation \citep{ChKur06},
while mostly axisymmetric fields with a mixture of
the first few multipoles are formed
in the case of a differentially rotating star\citep{DoStBr06}.
The results may not directly apply to pulsars or magnetars, 
but suggest that the magnetic field 
configuration of neutron stars may not be an ordered dipole.
If there are higher-order multipoles, these will also contribute 
to the radiation loss. The upper bounds on their surface magnetic fields
are rather loose. See \cite{Krol91} for a discussion of 
the magnetic fields of millisecond pulsars. 
The magnetic field strength $B_{lm}$ relevant to
the multipole moment of order $(l,m)$ is limited to
$ B_{lm} \le B_{s} /(m R_{s}\Omega /c)^{l-1} $,
where $\Omega=2\pi/P$ is angular velocity, and 
the radiation of each multipole 
$ L_{lm} \sim c ( m R_{s} \Omega /c) ^{2l+2} $
$ (B_{lm} R_{s} ) ^{2} $
is assumed to be smaller than that of a dipole.
Thus, a model with complex magnetic configuration
at surface $ B_{lm} \ge B_{s} ~( l > 1 ) $
is allowed because of the small factor $ R_{s} \Omega /c \ll 1 $
for observed stars.

  Some proto-neutron stars are conjectured to be born in hypothetical 
extreme state of rapid rotation $P \approx 1 $ ms
with an ultra strong magnetic field $B_{s} \approx 10^{15}$G.
Is there any remaining evidence of this stage? The proper motion 
can possibly be used as a probe.
Several kick mechanisms operative at the core bounce of 
a supernova explosion have been proposed to date:
anisotropic emissions of neutrinos (e.g., \cite{ArLa99,FrKu06}),
hydrodynamical waves (e.g., \cite{Scetal06}),
and MHD effects (e.g., \cite{SaKoYa08}).
These mechanisms operate on a dynamical timescale of the order of 
milliseconds or the cooling timescale of $\sim10$ s.
If the strong magnetic fields are generated on a longer timescale, 
some natal kick mechanisms involved the magnetic-field-driven 
anisotropy do not work effectively.
Recoil driven by electromagnetic radiation, which 
is operative on a longer spindown timescale of
$\sim 10^3 (B/10^{15} {\rm G})^{-2} (P_i/1{\rm ms})^{2}$ s, 
has been proposed as a post-natal kick mechanism\citep{HaTa75}
(see also \cite{LaChCo01} for the corrected expression).
In their model, an oblique dipole moment displaced 
by a distance $s$ from the stellar center rotates.
This causes the radiation of higher order multipoles,
whose superposition is generally asymmetric 
in the spin direction, leading to the kick velocity.
In the off-center model, the quadrupole field $B_{2}$
of order  
$B_{2} \sim (s/R_{s}) \times B_{1} \sim B_{1} $ is involved.
It is interesting to study the case where $ B_{2}  \gg   B_{1} $, 
because the constraint of the higher order component
by the radiation is very weak,  for example,   
$B_{2} \sim (c/( R_{s}\Omega)) \times B_{1} \gg   B_{1} $.
In this paper, we revisit the kick velocity induced by 
electromagnetic radiation from a magnetized rotating star with
both dipole and quadrupole fields, in which 
a larger quadrupole field $ B_{2}  \gg   B_{1} $
at the surface is allowed.

  This paper is organized as follows. In Section 2, we 
present the radiation from rotating dipole and quadrupole 
moments in vacuum. 
Their field strength and inclination angle with respect to the 
spin axis are arbitrary. 
We also compare our model with the off-centered dipole model.
In Section 3, we evaluate the maximum kick velocity 
as a recoil of momentum radiation.
Section 4 presents our conclusions.

 \section{MODEL}
 \subsection{ Electromagnetic Fields }
  We consider electromagnetic fields outside a 
rotating object with angular frequency $\Omega$; 
the object has a magnetic dipole and quadrupole moments.
The dipole moment is denoted by $\mu$, and the direction 
is inclined from the spin axis by $\chi_{1} $.
Quadrupole moment is denoted by $Q$ and 
the inclination angle of the symmetric axis 
is $\chi_{2} $ from the spin axis. 
The electromagnetic fields outside the rotating magnetized
object are described by the magnetic mupltipoles
of order $ l=1,2, |m| \le l$, for which  
$ E_r = {\bf E} \cdot {\bf r}=0$\citep{Jackson}.
The explicit components produced by the rotating dipole moment
are given by
\begin{equation}
B_{r}=\frac{2\mu }{r^{3}} 
\left[
 P^{0}_{1} (\chi_{1}) P^{0}_{1} (\theta )
+P^{1}_{1} (\chi_{1}) P^{1}_{1} (\theta ) S_{1}(\xi) e^{i\lambda } 
\right],
\label{eq1.br}
\end{equation}
\begin{equation}
B_{\theta }= - \frac{\mu }{r^{3}}
\left[
  P^{0}_{1} (\chi_{1}) P^{\prime 0}_{~1} (\theta )
+ P^{1}_{1} (\chi_{1}) P^{\prime 1}_{~1} (\theta )
  S_{2}(\xi) e^{i\lambda }
\right],
\label{eq1.bt}
\end{equation}
\begin{equation}
B_{\phi }=-\frac{i\mu }{r^{3}} P^{1}_{1} (\chi_{1})
S_{2}(\xi)e^{i\lambda },
\label{eq1.bp}
\end{equation}
\begin{equation}	
E_{\theta }=-\frac{\mu \Omega }{cr^{2}} P^{1}_{1} (\chi_{1})
S_{1}(\xi)e^{i\lambda },
\label{eq1.et}
\end{equation}
\begin{equation}
E_{\phi }=-\frac{i\mu \Omega }{cr^{2}} 
P^{1}_{1}(\chi_{1}) P^{\prime 1} _{~1}  (\theta )
S_{1}(\xi) e^{i\lambda },
\label{eq1.ep}
\end{equation}
where $\lambda =\phi -\Omega (t-r/c) $.
The function $P^{m} _{~l} (x)$ is the associated Legendre function and 
the prime denotes the derivative with respect to $x$.
The function $S_{n}$, which is derived from the spherical 
Hankel function, is a polynomial of $\xi= \Omega r/c$, and 
is explicitly written as
\begin{equation}
S_{1}(\xi)=1-i\xi,
\end{equation}
\begin{equation}
S_{2}(\xi)=1-i\xi-\xi^{2}.
\end{equation}
We here use convenient complex expressions in 
eqs. (\ref{eq1.br})-(\ref{eq1.ep}) and the 
actual fields are the real part. 
Near the stellar surface $ R_{s} \le r \ll c/\Omega$,
magnetic field for eqs.(\ref{eq1.br})-(\ref{eq1.bp})
at the phase $ e^{i\lambda } =1$
reduces to 
\begin{equation}
{\bf B} = \frac{ 2\mu }{r^{3}} \cos(\theta -\chi_{1}) {\bf e}_{r}
        + \frac{\mu }{r^{3}}   \sin(\theta -\chi_{1}) {\bf e}_{\theta }
        = - \nabla \left( \frac{ \mu }{r^{2}} \cos(\theta -\chi_{1})
                    \right).
\end{equation}
It is clear that the field near the origin represents
a magnetic dipole inclined by the angle $\chi_{1}$, which 
rotates in the azimuthal direction with $ \phi =\Omega t $.

  The electromagnetic fields for a rotating magnetic quadrupole
are similarly described by 
\begin{equation}
B_{r}=
  \frac{Q}{8r^{4}}\left[
  12 P^{0}_{2}(\chi_{2}) P^{0}_{2} (\theta )
  +4 P^{1}_{2}(\chi_{2}) P^{1}_{2} (\theta ) 
  S_{3}(\xi) e^{i\lambda _{2}}
  +P^{2}_{2}(\chi_{2}) P^{2}_{2} (\theta ) 
  S_{3}(2\xi)e^{2i\lambda _{2} }
\right],
\label{eq2.br}
\end{equation}
\begin{equation}
B_{\theta }=
  - \frac{Q}{24r^{4}} \left[
    12 P^{0}_{2}(\chi_{2}) P ^{\prime 0} _{~2} (\theta )
    +4 P^{1}_{2}(\chi_{2}) P^{\prime 1}_{~2} (\theta ) 
    S_{4}(\xi)e^{i\lambda _{2} }
    +P^{2}_{2}(\chi_{2}) P^{\prime 2}_{~2} (\theta )
     S_{4}(2\xi) e^{2i\lambda _{2} }
\right],
\label{eq2.bt}
\end{equation}
\begin{equation}
B_{\phi }=
 - \frac{iQ}{4r^{4}}\left[
   2 P^{1}_{2}(\chi_{2}) \cos \theta 
    S_{4}(\xi) e^{i\lambda _{2} }
  + P^{2}_{2}(\chi_{2})\sin\theta
   S_{4}(2\xi) e^{2i\lambda _{2} }
\right],
\label{eq2.bp}
\end{equation}
\begin{equation}
E_{\theta }=-\frac{Q\Omega}{4cr^{3}}\left[
   P^{1}_{2}(\chi_{2}) \cos \theta 
     S_{3}(\xi) e^{i\lambda _{2} }
  + P^{2}_{2}(\chi_{2}) \sin\theta
    S_{3}(2\xi) e^{2i\lambda _{2} }
\right],
\label{eq2.et}
\end{equation}
\begin{equation}
E_{\phi }=-\frac{iQ\Omega}{24cr^{3}}\left[
     2 P^{1}_{2} (\chi_{2}) P^{\prime 1}_{~2} (\theta ) 
        S_{3}(\xi)e^{i\lambda _{2} }
    +  P^{2}_{2}(\chi_{2}) P^{\prime 2}_{~2} (\theta )
       S_{3}(2\xi) e^{2i\lambda _{2} }
\right],
\label{eq2.ep}
\end{equation}
where 
\begin{equation}
S_{3}(\xi)=1-i\xi-\frac{1}{3}\xi^{2}, 
\end{equation}
\begin{equation}
S_{4}(\xi)=1-i\xi-\frac{1}{2}\xi^{2}+\frac{i}{6}\xi^{3}.
\end{equation}
The phase in eqs. (\ref{eq2.br})-(\ref{eq2.ep})
is shifted by $\lambda _{2} = \lambda +\delta$,
because the meridian plane in which the symmetric axis of 
the quadrupole is located may differ by the azimuthal angle
$\delta$ from that of the dipole.
The near-field of eqs.(\ref{eq2.br})-(\ref{eq2.bp}) 
for $ R_{s} \le r \ll c/\Omega$
at the phase $ e^{i\lambda _{2} } =1$ is 
\begin{equation}
{\bf B} = \frac{ 3Q }{2r^{4}} P^{0}_{2}(\theta -\chi_{2}){\bf e}_{r}
        - \frac{  Q }{2r^{4}} P^{\prime 0}_{~2}(\theta -\chi_{2})
         {\bf e}_{\theta } 
        = - \nabla \left( \frac{ Q }{2 r^{3}} 
          P^{0}_{2}(\theta -\chi_{2})
                    \right).
\end{equation}
Thus, the magnetic fields given by eqs. (\ref{eq2.br})-(\ref{eq2.bp})
are those of a rotating quadrupole, whose inclination angle 
is $\chi_{2}$.  

\begin{figure}[ht]
\begin{minipage}{0.45\linewidth}
  \includegraphics[scale=0.5]{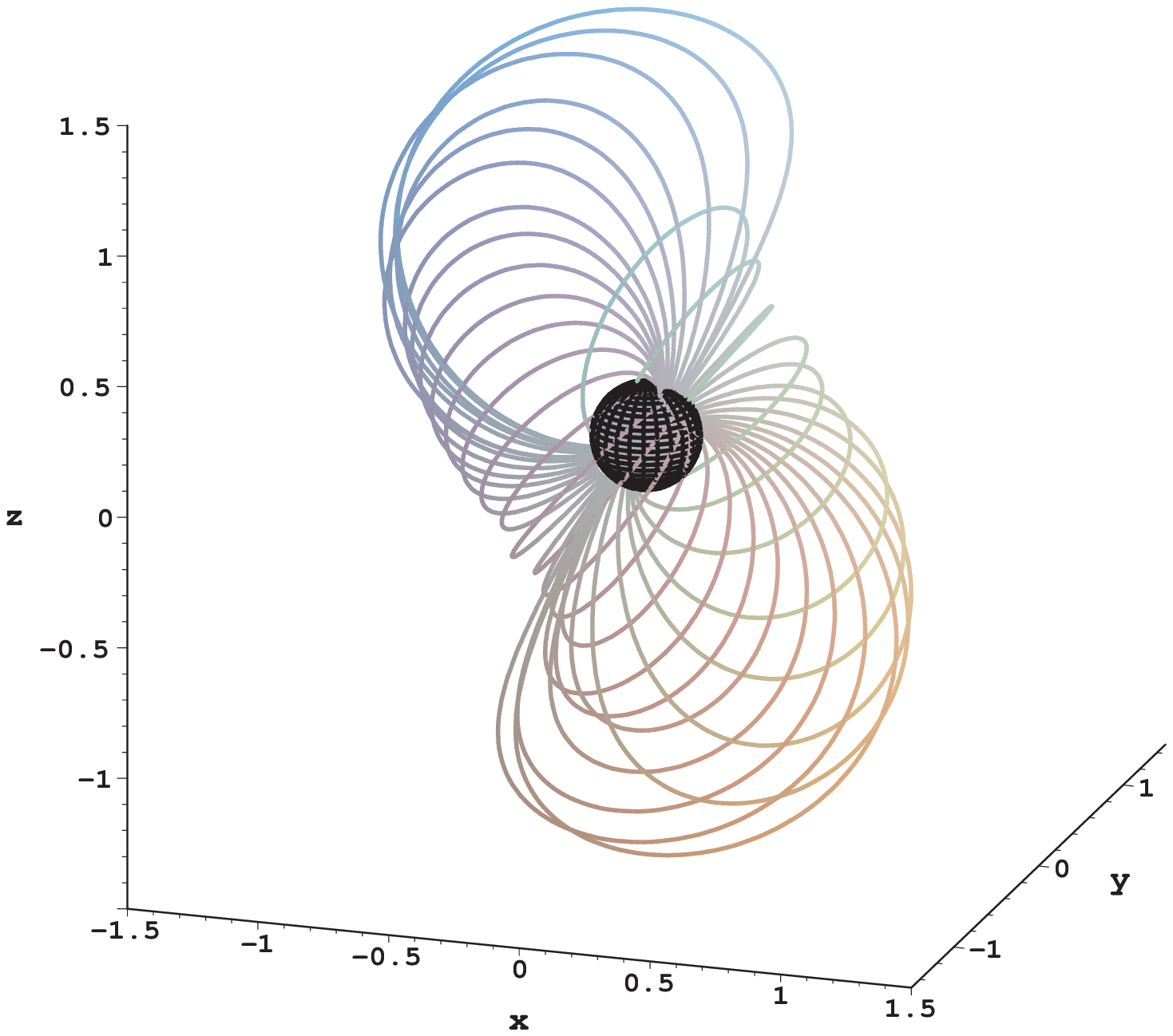}
\end{minipage}
\begin{minipage}{0.45\linewidth}
  \includegraphics[scale=0.5]{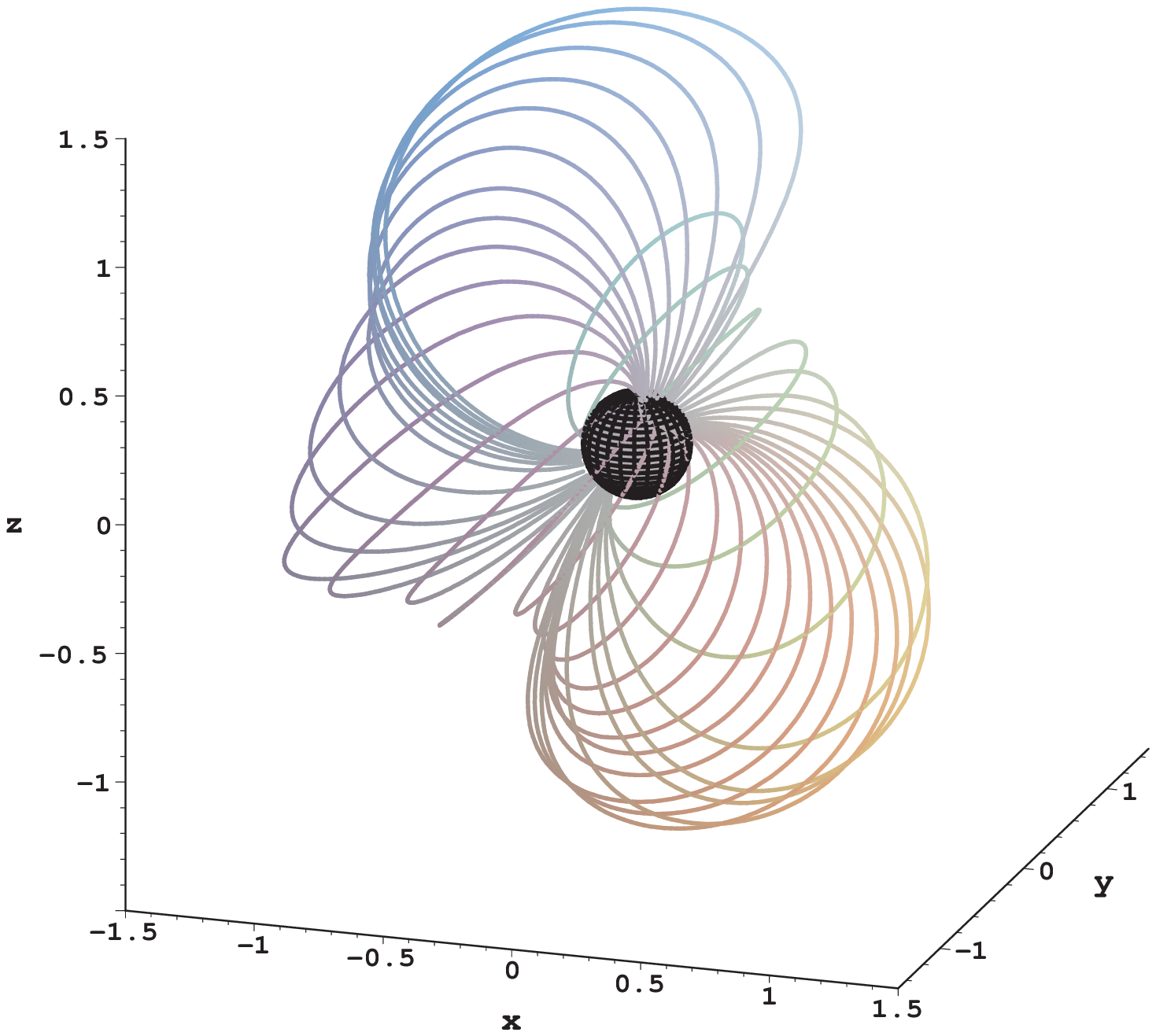}
\end{minipage}
\caption{ Closed magnetic field lines
in the cases of pure dipole (left)
and dipole plus quadrupole (right).
The magnetic axis is inclined by angle 
$\chi_{1}= \chi_{2}= \pi/4$ from the
spin axis $z$, and the azimuthal angle between 
moments is $ \delta =\pi/2$.
A sphere of radius $0.2 c/\Omega$ at the origin is also 
shown. Distance is scaled by $c/\Omega$.
\label{fig1}
}
\end{figure}

 We compare the combination of dipole and quadrupole fields
with the case of a pure dipole. A snapshot of almost-closed magnetic 
field lines near the light cylinder is shown in Fig. \ref{fig1}.
Both inclination angles are the same $\chi_{1}= \chi_{2}= \pi/4$, 
but the meridian planes are perpendicular, that is, $\delta = \pi/2$.
Field strength is set as $ Q = 0.2 \mu c/ \Omega $.
It is clear that the quadrupole field is added to the dipole one.
The quadrupole field increases more rapidly 
with the decrease of the radius $r$, and dominates 
for $ r < r_q$ $\approx 0.2 c/\Omega $ for the model parameter, 
since $B_{1} \sim \mu/r^3$ and $B_{2} \sim Q/r^4$.

  \subsection{ Radiation }
  The radiation energy per unit time is obtained by integrating 
the time-averaged Poynting flux over the solid angle 
at the wave zone $  r \gg c/ \Omega $.
The luminosity for a combination of electromagnetic fields described by 
eqs. (\ref{eq1.br})-(\ref{eq1.ep}) and  
eqs. (\ref{eq2.br})-(\ref{eq2.ep}) is given by
\begin{equation}
L= \int \frac{c}{4\pi} 
\overline{({\bf E} \times  {\bf B} )} 
\cdot {\bf e}_{r}  r^2 \sin  \theta d \theta d \phi
=
\frac{2 \mu^{2}\Omega ^{4} }{3c^3} \sin ^{2}\chi_{1}
+\frac{ Q^{2}\Omega ^{6}}{160c^5} \sin ^{2}2\chi_{2} 
+\frac{2 Q^{2}\Omega ^{6}}{5c^5} \sin ^{4}\chi_{2} .
\label{Erad_DQ.eqn}
\end{equation}
The luminosity is the sum of the contributions from 
multipole radiation. Our model consists of three components, 
the magnetic dipole radiation $M_{1,1}$ specified by spherical harmonics 
index $(l,m)=(1,1)$, and the quadrupole radiation $M_{2,1}$ and $M_{2,2}$.
They correspond to the first, second and third terms
in eq. (\ref{Erad_DQ.eqn}).
The third term is larger than the second term roughly
by a factor $m^6=2^6$, which comes from the frequency of time variation. 

  The linear momentum radiated per unit time in the direction $z$ 
is similarly calculated as
\begin{equation}
F = \int \frac{1}{4\pi}
\overline{({\bf E} \times  {\bf B} )} 
\cdot {\bf e}_{z}  r^2 \sin  \theta d \theta d \phi
=
\frac{\mu Q\Omega ^{5}}{20 c^5}
\sin \chi_{1} \sin 2\chi_{2} \sin \delta.
\label{Prad_DQ.eqn}
\end{equation}
The net flux arises from the interference of two multipoles, 
namely, the magnetic dipole $M_{1,1}$ and the quadrupole $M_{2,1}$.
The angle $\chi_{l} $ governs
the radiation strength of each multiple $l$, 
while the angle $\delta$ governs the interference.
The most efficient configuration is realized when the two magnetic 
multipole moments are orthogonal, $ \delta =\pi/2 $.
On the other hand, when both of the multipole moments lie in the same 
meridian plane (i.e., $ \delta =0 $), the net linear momentum vanishes.
This property can be understood from the fact that
radiative electromagnetic fields in vacuum are 
expressed by the spherical Hankel function $h_l$ 
and the asymptotic form for $\xi =\Omega r/c \gg 1$
is $ h_{l} \sim \exp[ i(\xi -l\pi/2 )]/r$
for the multipole $l$. There is a phase shift $ \pi/2 $ 
between dipole and quadrupole fields, and this shift is important
in the wave interference.

  \subsection{ Comparison }
 
  We compare our result with the off-center dipole model
\citep{HaTa75, LaChCo01}.
The rates of energy and linear momentum are written 
in term of the magnetic dipole moment
$ (\mu _{R},\mu _{\phi},\mu _{z} )$
in cylindrical coordinate and distance $s$ from the spin axis  
as follows:
\begin{equation}
L= \frac{2 \Omega ^{4} }{3c^3} 
\left(
\mu^{2} _{R} + \mu^{2} _{\phi} 
\right)
+\frac{ 4 \Omega ^{6}}{15c^5} s^{2} \mu^{2} _{z}.
\label{Erad_OC.eqn}
\end{equation}
The first term is the magnetic dipole radiation $M_{1,1}$.
Correspondence to our expression is clear by replacing 
$\mu^{2} _{R} + \mu^{2} _{\phi} = \mu^{2}  \sin^2 \chi_{1} $.
The second term is derived from the sum of
electric dipole radiation $E_{1,1}$
and magnetic quadrupole radiation $M_{2,1}$. 
Their contributions are 
$ \Omega ^{6} s^{2} \mu^{2} _{z} /(6c^5)$ by $E_{1,1}$ 
and 
$ \Omega ^{6} s^{2} \mu^{2} _{z} /(10c^5)$ by $M_{2,1}$, 
respectively. The parameter in the off-center dipole model
corresponds to $Q \sin 2 \chi_2 = 4 s \mu_{z}$
except for a complex phase factor. There is a constraint on
the quadrupole moment $Q$ as 
$Q \sin 2 \chi_2 \le 4 \mu R_{s} \cos \chi_1 $,
since $ s \le R_{s}$.  In our model, it is possible to consider 
the case of $ Q \gg \mu R_{s} $ in magnitude.

  The linear momentum in the off-center dipole model
is evaluated as\citep{LaChCo01}
\begin{equation}
F=\frac{8 \Omega ^{5} s \mu _{\phi} \mu _{z} }{15c^5}.
\label{Prad_OC.eqn}
\end{equation}
Net linear momentum flux arises from two
types of interference.  
One is between magnetic dipole radiation $M_{1,1}$ and 
electric dipole radiation $E_{1,1}$.
The other is 
between magnetic dipole radiation $M_{1,1}$ and 
magnetic quadrupole radiation $M_{2,1}$.
These contributions are expressed by 
$ \Omega ^{5} s \mu _{\phi} \mu _{z} /(3c^5)$ and 
$ \Omega ^{5} s \mu _{\phi} \mu _{z} /(5c^5)$, respectively.
The latter reduces to eq. (\ref{Prad_DQ.eqn})
if $s \mu_{z} = Q \sin 2 \chi_2 /4$ and
$ \mu _{\phi} =\mu \sin \chi_{1} \sin \delta$.

  Although there is a slight difference in the radiative components 
between the off-center dipole and dipole-quadrupole models,
both formulae for eqs. (\ref{Erad_DQ.eqn}),(\ref{Prad_DQ.eqn})
and eqs. (\ref{Erad_OC.eqn}),(\ref{Prad_OC.eqn})
are parameterized as
\begin{equation}
L=   \alpha \frac{\mu ^2 \Omega ^4 }{c^3}
   + \beta \frac{Q^2 \Omega ^6 }{c^5},
\label{Erad.eqn}
\end{equation}
\begin{equation}
F=  \frac{\gamma}{10} \frac{\mu  Q \Omega ^5}{c^5},
\label{Prad.eqn}
\end{equation}
where $\alpha$, $\beta$ and $\gamma$ are
dimensionless numbers that depend on only the geometrical configuration.
The typical values are listed in Table \ref{tab:comparison} for the simple assumption that $ \sin \chi_{l}, \sin \delta \to 1/\sqrt{2} $,
that is, the directional average of $ \langle  \sin^2 \chi_{l} \rangle $
$ =\langle  \sin^2 \delta \rangle =1/2$.
It is clear that the coefficient $\beta$ in our model is 
considerably larger than that in the off-center model.
This comes from the radiation of $m=2$.

\begin{table}
\begin{center}
\caption{ Comparison of models.}
\begin{tabular}{c c c c c c}
\hline \hline
Model &  Multipole &
 $\alpha$   &  $\beta$ 
 & $\gamma$  & $\gamma/(\alpha \beta )^{1/2} $  \\
\hline
 Off-center dipole & $M_{1,1},M_{2,1},E_{1,1}$ &
0.33 & 0.83 $\times 10^{-2}$ & 0.47 & 
 9.0
\\
 Dipole-quadrupole & $M_{1,1},M_{2,1},M_{2,2}$ &
0.33 & 0.10  & 0.18 & 0.97
\\
\hline \hline
\end{tabular}
\label{tab:comparison}
\end{center}
\end{table}

  \section{ EVOLUTION }
  We next consider the evolution of spin and kinetic velocity.
The angular velocity $\Omega (t)$ is determined by equating 
the loss rate of rotational energy with the luminosity $L$ 
in eq. (\ref{Erad.eqn}), and the velocity $V(t)$ is determined from the
momentum emission $F$ in eq. (\ref{Prad.eqn}).
In terms of the mass $M$ and inertial moment $I$, we have
\begin{equation}
I \Omega {\dot \Omega}
 =  -\alpha \frac{\mu ^2 \Omega ^4 }{c^3}
    - \beta \frac{Q^2 \Omega ^6 }{c^5},
\label{evl_spin}
\end{equation}
\begin{equation}
M {\dot V} =
  -\frac{\gamma}{10} \frac{\mu  Q \Omega ^5}{c^5} .
\label{evl_vel}
\end{equation}
By using the approximation $ I = 2M R_{s} ^2/5$, where
$R_{s}$ is the stellar radius, the magnitude of the
velocity gained from the initial angular velocity $\Omega_{i}$
is given by
\begin{equation}
\Delta V
 = \frac{ \gamma Q R_{s} ^2}{ 25 \mu  } \int _{\Omega_{0}}  ^{\Omega_{i}}  
\frac { \Omega ^2 }
{ \alpha c^2 + \beta ( Q /\mu )^2 \Omega ^2 }
d\Omega 
\le \Delta V_{*}
 \equiv \frac{ \gamma }{ 25 c ( \alpha \beta)^{1/2} } 
   \left( \Omega _{i} R_{s} \right) ^2
   X^{-2}\left[ X- \tan^{-1} X \right],
\end{equation}
where $ X \equiv ( \beta/\alpha )^{1/2} Q \Omega _{i} /\mu$
and the present angular velocity
$\Omega _{0} =0$ is used in the last inequality.
The function $\Delta V_{*} $ is determined by the ratio 
$ Q /( \mu R_s)$ of the two multipole moments 
for the fixed geometrical configuration
and the initial angular velocity $ \Omega _{i}$. 
Two limiting cases of $\Delta V_{*} $ are approximated as
\begin{equation}
\frac{\Delta V_{*} }{c}  
 \approx
\left\{
\begin{array}{ll}
 \frac{ \gamma}{75 \alpha}
 \left(   \frac{ Q }{ \mu R_s } \right)
 \left(  \frac{ \Omega _{i} R_s }{c} \right)^3 
&
 \mbox{for $0<X\ll 1$} 
\\
 \frac{ \gamma}{25 \beta}
\left(   \frac{ Q }{ \mu R_s } \right)^{-1}
 \left(  \frac{ \Omega _{i} R_s }{c} \right)
&
 \mbox{for $X\gg 1$}. 
\end{array}
\right.
\end{equation}

The value $ \Delta V_{*} $ increases as
the ratio $ Q /( \mu R_s)$ increases, while $ Q /( \mu R_s) \ll 1$, 
but begins to decrease for
$ Q /( \mu R_s) \to \infty$. Thus, it has a maximum 
with respect to the magnetic moment ratio:
\begin{equation}
 \frac{\Delta V_{*}}{c} 
  \approx 9.2 \times10^{-3}
  \frac{\gamma}{ (\alpha \beta)^{1/2}  }
 \left(  \frac{ \Omega _{i} R_s }{c} \right)^2 
~~~~
\mbox{ at }
 \frac{Q}{\mu R_s }  
  \approx 1.5 
  \left( \frac{ \alpha }{\beta } \right)^{1/2} 
  \left(  \frac{ \Omega _{i} R_s }{c} \right)^{-1}.
\label{opt.cond}
\end{equation}
The magnetic moment ratio at the maximum means that
the quadrupole field $B_{2} \sim Q/R_s ^4$ is
stronger than the dipole field $B_{1} \sim \mu/R_s ^3$ at the surface. 
The energy loss rate $ L_{l} $
of each multipole is approximately the
same at the beginning, 
$ L_{2} = \beta Q^2 \Omega_{i} ^6 / c^5$
$ \approx $
$  2.3 \times \alpha \mu ^2 \Omega_{i} ^4 / c^3 $
$  = 2.3 L_{1} $, but the contribution of $ L_{2} $ 
and becomes smaller as $\Omega$ is decreased.
The velocity using the canonical values is evaluated as
\begin{equation}
\Delta V_{*}
\approx 120 
\left(\frac{P_{i} }{ 1 {\rm ms} } \right)^{-2}
\times
\frac{\gamma}{ (\alpha \beta)^{1/2} } 
~{\rm km~s}^{-1}
\label{terminal}
\end{equation}
For the off-center dipole model, 
$V_{*} \sim 10^3 (P_{i}/1{\rm ms})^{-2}$ km s$^{-1}$
is allowed for an initially rapid rotator,
the initial period $P_i =1$ ms,
using typical values given in Table \ref{tab:comparison}. 
On the other hand, the typical value 
is small, $V_{*} \sim 10^2 (P_{i}/1{\rm ms})^{-2}$ km s$^{-1}$,
for the dipole-quadrupole model. The difference comes from 
the presence of radiation of $m=2$, which causes efficient 
energy loss, as discussed in the previous section.
Nevertheless, extremely high velocity is possible for
a specific configuration even in the present model.
Small $\beta$ corresponds to high velocity.
For small $\chi _{2}$ in eq. (\ref{Erad_DQ.eqn}),
we have $\beta  = \sin ^{2} 2\chi _{2} /160$. 
Because $\alpha = 2 \sin ^{2} \chi _{1} /3$, 
$\gamma = \sin  \chi _{1}  \sin  2\chi _{2} /2$ 
for $\sin \delta =1$; the combination of parameters reduces to 
$\gamma /(\alpha \beta)^{1/2}  =7.7$.
The resultant kick velocity increases up to 
$ \sim 930 (P_{i}/1{\rm ms})^{-2}$ km s$^{-1}$.
This optimal case corresponds to the magnetic configuration 
with an inclined dipole and a nearly axially symmetric 
quadrupole. The ratio of the moments is
$ Q /( \mu R_s) \sim$
$74 (\sin \chi _{1}/ \sin \chi _{2} )( P_{i}/ 1 {\rm ms} )$.

\begin{figure}[ht]
\begin{center}
 \includegraphics[scale=0.75]{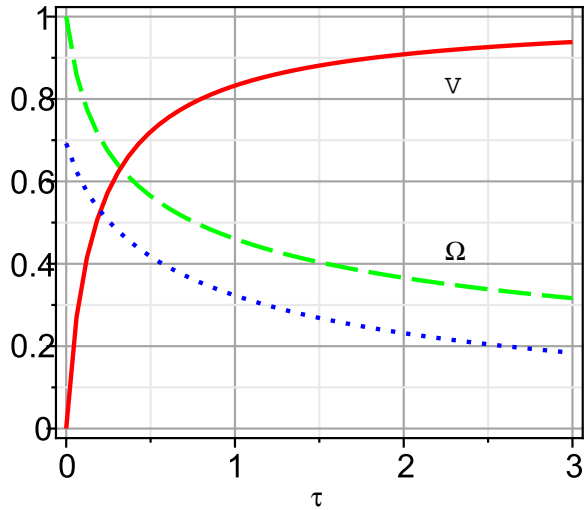}
\caption{ Time evolution of velocity, spin and
energy loss rates as a function of dimensionless time $\tau =t/t_*$.
Solid line represents the linear velocity normalized by 
the terminal one $\Delta V_{*}$, 
and dashed line the angular velocity normalized by 
initial one $\Omega _{i}$.
Dotted line is the radiation loss ratio
$L_{2}/(L_{1}+L_{2})$.
\label{fig2}
}
\end{center}
\end{figure}

Time evolution of spin and velocity is calculated for
the optimized relation (\ref{opt.cond}). Once the
quadrupole field strength is fixed, the
evolution of $\Omega(t)$ in eq.(\ref{evl_spin})
is scaled by characteristic time $t_{*}$
of the dipole radiation loss for the initial
angular velocity $\Omega _{i}=2\pi/  P_{i} $:
\begin{equation}
 t_{*} = \frac{ I c^3 }{ \alpha \mu^2 \Omega _{i} ^{2}  }
  \approx 0.8 
  \left(  \frac{ \alpha }{ 0.33 } \right)^{-1} 
  \left(  \frac{ P_{i} }{ 1 {\rm ms} } \right)^2 
  \left(  \frac{ \mu }{ 10^{31} {\rm G cm}^3 } \right)^{-2}
~~\mbox{ yr },
%
\end{equation}
where magnetic dipole field at the surface
is chosen as $ B_{1} \sim 10^{13}$ G. 
Figure \ref{fig2} shows the evolution of
 $\Omega (t) /\Omega _{i}$
as a function of $\tau = t/t_{*}$.
The ratio of quadrupole to the total energy loss rate,
$ L_{2}/( L_{1} + L_{2} )$ is also plotted.
The ratio at $t=0$ is approximately $0.7$
because of $L_{2} \approx 2.3 L_{1}$,
but monotonically decreases.
At $t=t_{*}$, the angular velocity becomes
$\Omega \sim 0.5 \Omega _{i}$ and
the contribution of quadrupole radiation also
decreases as
$L_{2} \sim 0.5 L_{1}$.
The velocity $V(t)$  normalized by
the terminal one (eq.(\ref{terminal})) is also
shown in Fig.\ref{fig2}.
The magnitude attains to almost terminal value,
$V \sim 0.8 \Delta V_{*}$ before $t=t_{*}$.

  \section{CONCLUSIONS}

  Magnetic field strength itself is critical in most kick mechanisms. For example, $ B_{s} > 10^{15}$ G 
at the surface is required in asymmetric neutrino 
emission (e.g., \cite{ArLa99}), as well as in asymmetric magnetized core 
collapse (e.g., \cite{SaKoYa08}).
The resultant velocity increases with the field strength because
the asymmetry arises from the magnetic field.
Magnetars are therefore expected to have high velocity if 
one of these mechanisms is operative.
Recent observations do not support the high velocity.
Rather, the upper limit of the transverse velocity $v_{\bot}$
has been reported, although there is uncertainty in the value. 
For example, 
$v_{\bot} \sim 210$ km s$^{-1}$ for AXP XTEJ1810-197\citep{Helfandetal07},
$v_{\bot} < 1300$  km s$^{-1}$ for SGR 1900+14 
\citep{Kaplanetal09, Lucaetal09}
and $v_{\bot} < 930$  km s$^{-1}$ for AXP 1E2259+586\citep{Kaplanetal09}.
For fast moving pulsars, 
PSR2224+45  ($v_{\bot} > 800$  km s$^{-1}$ \cite{CoRoLu93}) and
B1508+55    ($v_{\bot} \sim 1000$  km s$^{-1}$ \cite{Chatterjee05})
have been reported. 
These magnetic fields are quite ordinary,
$B_{s} =2.7\times 10^{12}$G and $2.0\times 10^{12}$G, respectively.
Thus, there is no clear correlation between the field strength 
and the velocity in the present sample.

  The electromagnetic rocket mechanism considered in \cite{HaTa75} 
and in this paper does not depend on field strength if the spin evolution is determined from the radiation loss.
In our model, the ratio of dipole and quadrupole moments is important.
The condition for high velocity is that the quadrupole
field is large enough in magnitude for the radiation loss 
to be of the same order as the dipole field.  The velocity also depends 
on the geometrical configuration of the multipole moments,
that is, each inclination angle from the spin axis and the angle between 
the axes of symmetry of the moment.
Assuming that the directions of moments are random, and 
that they are equally likely to be oriented in any direction,
it is found that the mean velocity with respect to the configuration
is not so large, $ \sim 120 (P_{i}/1{\rm ms})^{-2}$  km s$^{-1}$, for
the optimized dipole-quadrupole ratio.
The maximum velocity is realized for a specific configuration 
in which the inclination angle of the quadrupole moment is 
small, and the meridian plane in which the quadrupole moment lies 
is perpendicular to the plane of the dipole.
The velocity increases up to $ \sim 930 (P_{i}/1{\rm ms})^{-2}$  km s$^{-1}$.
This value is slightly smaller than the maximum observed velocity of a pulsar.

  The configuration is unknown, and is closely related to
the origin of the magnetic field, dynamo or fossil.
Nevertheless, \cite{BoUrBe06} reported interesting 
results within the mean-field dynamo theory. 
They argued that strong large-scale and weak small-scale fields are 
generated only in a star with a very short initial period, that is, the Rossy number is small,
and that the maximum strength decreases and
small-scale fields become dominant with the decrease of the 
initial period.
Thus, magnetars may have an ordered dipole with a strong field, 
while some pulsars may have rather irregular fields with 
higher multipoles.
Through the superposition of higher multipoles, 
pulsars in general come to have a larger 
radiation recoil velocity than magnetars.

  Finally, if the kick velocity of pulsars and magnetars is governed
by the same mechanism, 
it either should not simply depend on
magnetic field, or should depend on only the configuration.
The latter possibility was explored here.
Present argument is recognized as the order of magnitude
level due to the rotating model in vacuum.
Further improvement of the magnetosphere will be
of importance to explore the idea.
  
\section*{Acknowledgements}
This work was supported in part by the Grant-in-Aid for
Scientific Research (No.21540271) from
the Japanese Ministry of Education, Culture, Sports,
Science and Technology.


\end{document}